# Sensitivity Comparison of Microwave-Frequency and Optical Fibre Interferometry Based on State-of-the-Art Components

Georgios Aias Karydis[(1)], Marco Fasano[(2)], Paola Parolari[(2)], Pierpaolo Boffi[(2)], Charis Mesaritakis [(3)], Adonis Bogris[(1)]

(1) Department of Informatics and Computer Engineering, University of West Attica, Aghiou Spiridonos, 12243, Egaleo, Athens, Greece (gakarydis@uniwa.gr, abogris@uniwa.gr )
(2) Politecnico di Milano, DEIB, via Ponzio 34/5, Milano, Italy
(3) Department of Biomedical Engineering, University of West Attica, Aghiou Spiridonos, 12243, Egaleo, Athens, Greece

**Abstract** *We compare the sensitivity of fibre interferometers to vibrations using microwave oscillators and state-of-the-art lasers over distances up to 70 km. High spectral purity lasers outperform only above 10 Hz, highlighting microwave advantages at low frequencies and the potential of hybrid microwave–optical interferometry systems.* 

## Introduction

Fibre optics sensing techniques that are incorporated in already installed cable networks have gained significant traction the last decade due to their multiple potential applications spanning from optical network surveillance, smart cities applications to geophysical and environmental monitoring [1-3]. The dominant way to implement fibre sensing is with the use of a stable laser oscillator either in reflectometry or interferometry configurations. The former approach has given rise to a large number of distributed acoustic sensing (DAS) implementations, such coherent optical time domain reflectometry (C-OTDR), phase-OTDR and optical frequency domain reflectometry (OFDR) [4,5]. The use of interferometry has been promoted in submarine cables for interrogating very long links mostly targeting geophysical monitoring in the deep ocean [5,6]. In all implementations, the sensitivity of the interrogator is determined by the capability of the technique in acquiring phase changes over the phase noise of the laser source. This is why fibre lasers with sub-kHz linewidth are employed in all commercially available DAS systems, whilst in long-haul implementations, even more sophisticated lasers, with sub-Hz linewidth, are preferred to expand the coherence length of the source to thousands of km [6].

Recently, the use of microwave sources in fibre interferometry and DAS experiments has been proposed [7-9]. The sensitivity of phase acquisition in the microwave domain is lower compared to optical phase measurements as phase changes are proportional to the interrogation frequency. Despite the much lower phase signal offered by the interrogation in the microwave domain, the use of microwave sources has critical advantages. On one hand, the low phase noise of state-of-the-art microwave oscillators guarantees better noise performance at low frequencies especially if geophysical monitoring is targeted [7,8]. On the other hand, DAS systems employing microwave carriers have a much higher saturation limit in large strain rate variations compared to typical DAS approaches which make them an interesting candidate for early warning systems [9]. Despite the parallel implementations of DAS and interferometry systems based on laser and microwave carriers in the literature, a systematic work comparing their sensitivity is missing with the exception of very recent theoretical paper focused on submarine cables [10].

In this work, we implement microwave frequency fibre interferometers (MFFI) and optical interferometers (OFI) with path lengths ranging from 300 m to 70 km, using state of the art commercially available oscillators, namely 10-GHz microwave generators and a fibre laser with100-Hz linewidth. We compare MFFI and OFI in terms of phase noise and sensitivity to well controlled stimuli generated by a fibre stretcher. The analysis clearly shows the superiority of OFI at frequencies above > 10 Hz in all studied paths and the trend of MFFI to provide equal or even better sensitivity at frequencies below 5 Hz. These results clearly demonstrate that MFFI and OFI are complementary modalities, and that their integration into a hybrid system enables a highly powerful interrogator with exceptional sensitivity and a broad dynamic range spanning sub-mHz to kHz acoustic sensing.

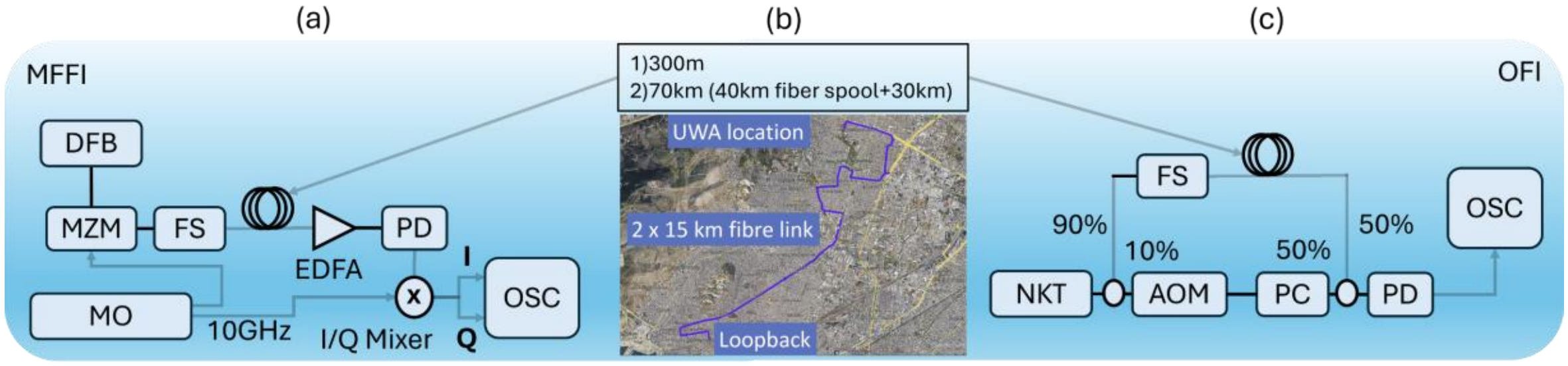

**Fig. 1:** (a) MFFI setup, (b) fibers used in the experiment, (c) OFI setup. MZM: Mach-Zehnder modulator, MO: Microwave oscillator, FS: Fibre Stretcher, DFB: Distributed Feedback Laser, PD: Photodiode, NKT: NKT Koheras Laser E15, AOM: Acousto-optic modulator, PC: Polarization Controller, OSC: Oscilloscope.

**Experimental setup**

MFFI and OFI implementations are depicted in fig. 1. The MFFI system consists of a distributed feedback (DFB) laser, an intensity Mach-Zehnder modulator (MZM), the fibre under test and a photodiode (PD). An Erbium-doped fibre amplifier (EDFA) before the PD compensates optical power losses when longer fibres are monitored. Two microwave oscillators have been tested, namely a Rohde & Schwarz SMA100B RF (RS) generator and a low-cost Phase Locked Loop (PLL) evaluation board [11]. The oscillators produce a sinusoidal signal applied on the MZM to modulate light at 10 GHz and to the local oscillator (LO) of an I/Q RF mixer that down-converts the I/Q components in the baseband. A standard 1550-nm DFB laser with 300 kHz linewidth and output power of 10 dBm is the light source for the optical path. In this configuration, the laser's phase noise plays a minimal role, as the signal is directly detected and the phase comparison takes place in the microwave domain. A fibre stretcher is included in the optical path to apply acoustic perturbations of frequencies as low as 5 Hz up-to 10 kHz. In this work, we construct two interferometers: one using a 300 m fibre spool, and another with a total path length of 70 km, consisting of a 40 km spool and an additional 30 km of deployed fibre in the metropolitan area of Athens, as shown in Fig. 1(b). Regarding the 70 km fibre, losses are 22.7dB in total which are compensated using an EDFA before photodetection. A ThorLabs 20 GHz PD was used in the receiving end of the optical path and the RF signal goes to the RF input of the I/Q RF mixer, which down-converts the LO 10-GHz frequency to DC in I and Q components. The I, Q components are measured by a Keysight DSOV084A oscilloscope, acquiring data at 1MSa/s for 50 seconds for each measurement. The phase is estimated through the I/Q components [10].

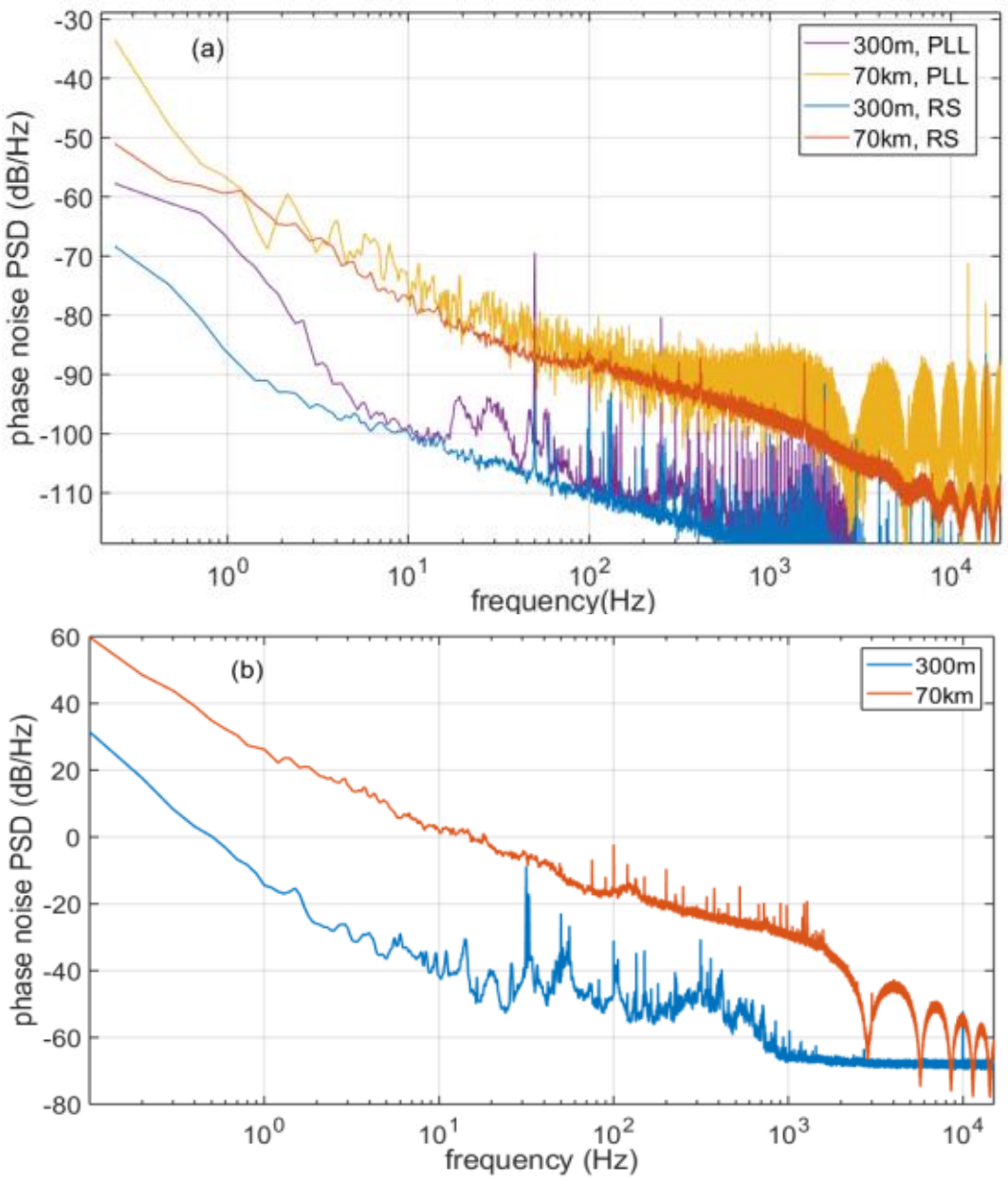

**Fig. 2:** Phase noise PSD for MFFI (a) and OFI (b) at 300 m (blue) and 70 km (red) interferometer lengths.

In the OFI case we measure the phase employing the same Mach-Zehnder delay interferometer but using optical coherent heterodyne detection (fig 1b). In this modality we use a NKT Koheras E15 laser with a linewidth of 100 Hz. The 15 dBm output power is split by a 90/10 1x2 coupler. The 90% is fed into the fibre under test while the 10% is sent into an acousto-optic modulator (AOM) which shifts the optical frequency by 40 MHz. A second coupler is used to combine the light from the two OFI arms, which is then detected by the receiving PD. After the AOM and before the second coupler a polarization controller (PC) is used to maximize the optical interference. The 40 MHz signal is acquired by the same oscilloscope and down-converted offline to extract the I, Q components and estimate the phase measured by the OFI. The PD, fibre spools, deployed 30 km long fibre and oscilloscope were common parts of the MFFI and OFI setups. We used a lower quality laser in MFFI in order to show its compatibility with low-cost laser sources. Moreover, for the OFI, due to the 40 MHz frequency shift applied by the AOM, the beating at 40 MHz in the PD forces the acquisition sampling rate at 100 MSa/s. 50 seconds of acquisition was followed in both cases for the phase noise power spectral density estimation.

**Results and Discussion**

In order to compare the two systems in terms of sensitivity, first we measured the phase noise power spectral density at 300 m and 70 km. The results are shown in fig. 2. The MFFI has a substantially lower phase noise PSD compared to OFI due to the lower frequency noise of the microwave sources at 10 GHz compared to that of the fibre laser counterpart. The high-quality RS generator outperforms the PLL in terms of the phase noise performance with a higher advantage at low frequencies (5 to 20 dB advantage). Nevertheless, the latter provides a satisfactory performance given its very low cost (< 600 euros). For a 300 m long interferometer, the MFFI exhibits -100 dB/Hz at 10 Hz, whilst OFI achieves a value of -40 dB/Hz which translates to a 60 dB noise advantage of MFFI over OFI. This noise advantage becomes around 80 dB for the 70 km long interferometer for the same frequency of 10 Hz when the RS oscillator is considered. The noise advantage is significant; however, it is also important to emphasize that OFI exhibits a remarkable signal advantage over its MFFI counterpart. This arises from the fact that the detected phase is directly proportional to the oscillator frequency. Consequently, OFI operating at 192 THz provides a gain factor of $192\,\text{THz}/10\,\text{GHz}$ relative to an MFFI system operating at 10 GHz in terms of the signal amplitude. This corresponds to an approximately 43 dB increase in the detected phase amplitude for the same physical effect, which translates to an 86 dB gain in the power spectrum. Therefore, despite the superior noise performance of MFFI, the substantial 86 dB signal gain offered by OFI must be taken into account to obtain a complete assessment of the sensitivity of the two systems.

In order to evaluate the sensitivity, we applied controlled stimuli to both MFFI and OFI interrogators by means of a fibre stretcher as depicted in fig. 1. We applied frequencies $f_{str}$ of 5, 10, 70, 130 and 190 Hz on the stretcher which lie within the acoustic bandwidth of interest for detecting anthropogenic or geophysical activities. 3 Volts of sinusoidal stimuli were applied on the driver of the stretcher which correspond to a maximum elongation of 9 μm. For this elongation, the maximum phase variation in the OFI is approximately 55 rads whilst in MFFI this is equal to 3 mrad. Then we calculated the carrier to noise ratio (C/N) defined as:

$$\frac{C}{N} = \frac{<\varphi_{str}^2>}{S_\varphi(f_{str})*1Hz} \qquad (1)$$

Where $\varphi_{str}^2$ is the phase signal power in rad$^2$ and which is 86 dB higher in the case of OFI, $S_\varphi(f_{str})$ is the value of the phase noise PSD at frequency $f_{str}$ which, as said, is evaluated from 5 to 190 Hz and 1 Hz is the integration bandwidth, where we consider that the phase noise PSD remains constant.

The results of this study are shown in fig. 3 for 300 m and 70 km lengths. For small interferometry lengths, a state-of-the-art fibre laser provides a 20 dB – 30 dB advantage in detectivity expressed in C/N when compared to a microwave generator. This advantage becomes higher as the frequency of the vibration increases. In contrast, for an interferometer of 70 km, the two systems have a lower gap in detectivity and they provide comparable performance at frequencies below 10 Hz. At 5 Hz they offer nearly the same C/N of 15 dB when the RS oscillator is used. A marginal 5 dB advantage for the OFI is observed when a low-cost PLL is used. This result clearly shows that for low frequency events, which usually correspond to geophysical or environmental effects, and for interferometer lengths exceeding 50 km, MFFI could complement OFI as a potentially more sensitive modality. We have also estimated the noise integrated in the bandwidth [5 Hz - 50 Hz] for the two cases assuming the RS oscillator for MFFI. The RMS phase noise is equal to $8\times10^{-4}$ rad for MFFI and 8.4 rad for OFI. This corresponds to a noise ratio of $20\log10(8.4/8\times10^4)$ = 80.4 dB between OFI and MFFI, yielding an overall advantage of only 5.6 dB for OFI within the bandwidth where the majority of anthropogenic and geophysical activity is concentrated. This advantage vanishes if frequencies lower than 5 Hz are of interest such as teleseismic earthquakes, ocean activity or tsunamis which exhibit strong content at sub-Hz frequencies [5, 6].

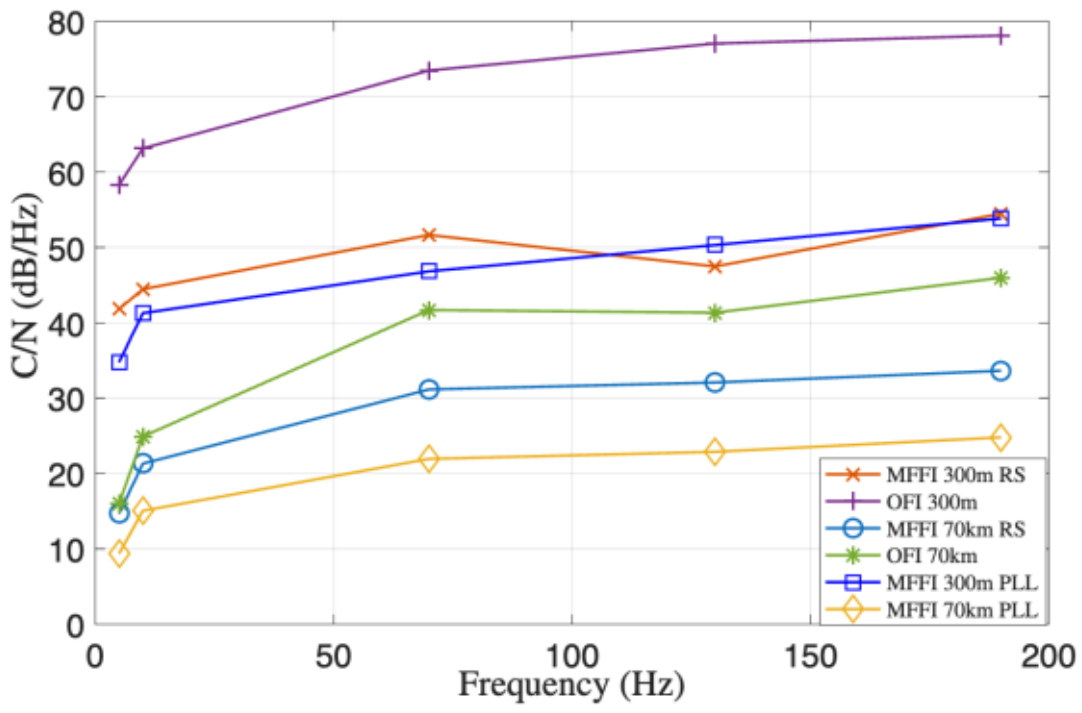

**Fig. 3:** C/N as a function of frequency in the acoustic noise bandwidth

## Conclusions

We presented a comparative analysis between OFI and MFFI interferometry schemes in terms of their detectivity in external vibrations within a frequency range from 5 Hz to 200 Hz and interferometer lengths of 300 m and 70 km. The analysis clearly shows that the two systems become comparable when frequencies below 10 Hz are investigated. The lower cost of MFFI compared to OFI makes it a promising technique for frequencies of interest below 5 Hz and opens the way for the development of hybrid OFI/MFFI systems for higher sensitivity and better dynamic range across a wide range of frequencies.

## Acknowledgements

This work has been supported by Horizon Europe project ECSTATIC under grant agreement 101189595.

## References

[1] Hang Wang, Yunfeng Chen, Rui Min, Yangkang Chen, "Urban DAS data processing and its preliminary application to city traffic monitoring," *Sensors*, vol. 22, no. 24, 2022. DOI: 10.3390/s22249976

[2] Christian Dorize, Sterenn Guerrier, Elie Awwad, Haïk Mardoyan, Jérémie Renaudier, "From coherent systems technology to advanced fiber sensing for smart network monitoring," *Journal of Lightwave Technology*, vol. 41, no. 4, pp. 1054-1063, 2023. DOI: 10.1109/JLT.2022.3221552

[3] Antonio Mecozzi, "Sensing with submarine optical cables," *Applied Physics Letters Photonics*, vol. 9, no. 7, 2024. DOI: 10.1063/5.0210825

[4] Zuyuan He, Qingwen Liu. "Optical fiber distributed acoustic sensors: A review," *Journal of Lightwave Technology*, vol. 39, no.12, pp. 3671-3686, 2021. DOI: 10.1109/JLT.2021.3059771

[5] Mikael Mazur, Nicolas K. Fontaine, Roland Ryf, Martin Karrenbach, Keith L. McLaughlin, Berry J. Sperry, Anuar G. Butler, Valey Kamalov, Lauren Dallachiesa, Ellsworth Burrows, David Winter, Haoshuo Chen, Jeewan Naik, Kishore Padmaraju, Ajay Mistry, David T. Neilson, "Submarine cable deep-ocean observation of mega-thrust earthquake and tsunami with 44,000 100-m spaced sensors," In *2025 European Conference on Optical Communications (ECOC)*, 2025. DOI: 10.48550/arXiv.2509.24813

[6] G. Marra, D. M. Fairweather, V. Kamalov, P. Gaynor, M. Cantono, S. Mulholland, B. Baptie, J. C. Castellanos, G. Vagenas, J.-O. Gaudron, J. Kronjäger, I. R. Hill, M. Schioppo, I. Barbeito Edreira, K. A. Burrows, C. Clivati, D. Calonico, A. Curtis, "Optical interferometry–based array of seafloor environmental sensors using a transoceanic submarine cable," *Science*, vol. 376, no. 6595, pp. 874-879, 2022. DOI: 10.1126/science.abo1939

[7] Adonis Bogris, Thomas Nikas, Christos Simos, Iraklis Simos, Konstantinos Lentas, Nikolaos S. Melis, Andreas Fichtner, Daniel Bowden, Krystyna Smolinski, Charis Mesaritakis, Ioannis Chochliouros, "Sensitive seismic sensors based on microwave frequency fiber interferometry in commercially deployed cables," *Scientific Reports*, vol. 12, 2022. DOI: 10.1038/s41598-022-18130-x

[8] A. Bogris, C. Simos, I. Simos, Y. Wang, A. Fichtner, S. Deligiannidis, N. S. Melis, C. Mesaritakis, "Microwave Frequency Fiber Interferometry in Submarine Deployed Telecommunication Cables," *presented at Optical Fiber Communications Conference and Exhibition (OFC)*, San Diego, USA, 2025, arXiv, 2025. DOI: 10.48550/arXiv.2504.05369

[9] Yan Ren, Pedro Vidal-Moreno, María R. Fernández-Ruiz, Sonia Martín-López, Luis Costa, Zhongwen Zhan, Miguel González-Herráez, "Microwave frequency OTDR for high strain-rate sensing." In *29th International Conference on Optical Fiber Sensors*. Vol. 13639. SPIE, 2025. DOI: 10.1117/12.3062910

[10] G. A. Karydis, M. Skontranis, C. Simos, I. Simos, T. Nikas, C. Mesaritakis and A. Bogris, "Per-Span Microwave-Frequency Fiber Interferometry for Amplified Transmission Links Employing High-Loss Loopbacks," Sensors, vol. 26, pp. 2551-2564, 2026, DOI: 10.3390/s26082551

[11] Analog devices, "ADF41513 26.5 GHz, Integer N/Fractional-N, PLL Synthesizer"